\documentclass[12pt]{article}
\usepackage{dcolumn}
\usepackage{graphicx}
\usepackage{epstopdf}
\usepackage{amsmath}
\usepackage{amsfonts}
\usepackage{amssymb}
\usepackage{color}
\usepackage{mathrsfs}

\newcommand{\be}{\begin{equation}}
\newcommand{\ee}{\end{equation}}
\newcommand{\bea}{\begin{eqnarray}}
\newcommand{\eea}{\end{eqnarray}}
\newcommand{\barr}{\begin{array}}
\newcommand{\earr}{\end{array}}

\def\beq{\begin{equation}}
\def\eeq{\end{equation}}
\def\be{\begin{equation}}
\def\ee{\end{equation}}
\def\bea{\begin{eqnarray}}
\def\eea{\end{eqnarray}}
\def\d{{\partial}}

\def\fnl{$f_{NL}\ $}
\def\taunl{$\tau_{\rm NL}\ $}
\def\mpl{M_{\rm Pl}}

\begin{document}


\setcounter{page}{1} \baselineskip=15.5pt \thispagestyle{empty}

\begin{flushright}
\end{flushright}
\vfil

\begin{center}

{\Large \bf A Naturally Large Four-Point Function 
\\[0.2cm]
in Single Field Inflation}
\\[0.7cm]
{\large Leonardo Senatore  and Matias Zaldarriaga}
\\[0.7cm]
{\normalsize { \sl School of Natural Sciences, Institute for Advanced Study, \\Olden Lane, 
Princeton, NJ 08540, USA}}\\
\vspace{.3cm}

\end{center}

\vspace{.8cm}

\hrule \vspace{0.3cm}
{\small  \noindent \textbf{Abstract} \\[0.3cm]
\noindent Non-Gaussianities of the primordial density perturbations have emerged as a very powerful possible signal to test the dynamics that drove the period of inflation. While in general the most sensitive observable  is the three-point function in this paper we show that there are technically natural inflationary models where the leading source of non-Gaussianity is the four-point function. Using the recently developed Effective Field Theory of Inflation, we are able to show that it is possible to impose an approximate parity symmetry and an approximate continuos shift symmetry on the inflaton fluctuations that allow, when the dispersion relation is of the form $\omega\sim c_s k$,  for a unique quartic operator, while approximately forbidding all the cubic ones. The resulting shape for the four-point function is unique. In the models where the dispersion relation is of the form $\omega\sim k^2/M$ a similar construction can be carried out and additional shapes are possible. 
}
 \vspace{0.3cm}
\hrule
\vfil



\section{Introduction}

In the last few years, it has become more and more clear that single-clock inflation, {\it i.e.} inflationary models where there is only one degree of freedom driving inflation, can produce large and detectable non-Gaussianities. While standard slow roll inflation cannot produce large non-Gaussianities~\cite{Maldacena:2002vr}, several models in single field inflation have been proposed that result in a detectable level of non-Gaussianities~\cite{Alishahiha:2004eh,ArkaniHamed:2003uz,Senatore:2004rj,Chen:2006nt,Cheung:2007st,Senatore:2009gt,Flauger:2009ab}. Furthermore thanks to the development of an Effective Field Theory for the Inflationary perturbations \cite{Cheung:2007st,Senatore:2009gt}, it has become clear that the Lagrangian for the fluctuations can incorporate large interactions without spoiling the naturalness  of the quasi de Sitter background. 

Non-Gaussianities represent a powerful signal about inflation for two reasons. First contrary to what happens for the two-point function, non-Gaussianities are sensitive to the interacting part of the inflaton Lagrangian which is clearly more interesting than the free-field part. Second, the symmetries of the inflationary spacetime force the two-point function of the cosmological fluctuations to be quasi scale invariant. This leaves freedom only to vary its amplitude and to add a small scale-dependence as we vary the inflationary models. This means that measurement of the inflationary two-point function reduces in practice to measuring only a couple of numbers. On the other hand, the same symmetries of the inflationary spacetime are not able to constrain as much correlation functions of higher order.  The possible signals in this case are continuous functions of several variables, very similar to the angular dependence of a scattering amplitude~\cite{Senatore:2009gt,Babich:2004gb}. This is clearly a much more powerful signal.

So far, the observational search for non-Gaussianity as originated from the period of inflation has concentrated on the three-point function. Several models predicted such a signal as the leading source of non-Gaussianity. The purpose of this paper is to show using the Effective Field Theory of Inflation~\cite{Cheung:2007st} that there are technically natural models of inflation where the level of the (connected) four-point function is much larger than the three-point function, and where the leading source for detection is indeed in the four-point function.  Using the Effective Field Theory of Inflation we show that it is possible to impose on the inflaton fluctuations both an approximate continuous shift symmetry and an approximate parity symmetry that, for models not close to the de Sitter limit, allow for a unique large quartic operator, $\dot\pi^4$, where $\pi$ represent the inflationary fluctuations. These symmetries approximately forbid all  cubic terms.  The resulting shape of the four-point function is unique. A similar construction is carried out for models near the de Sitter limit where more that one shape becomes possible.

\section{Effective Field Theory of Single-Clock Inflation\label{sec:single-field}}

In this section we briefly review the effective action for single-clock inflation. This effective
action was developed in \cite{Cheung:2007st,Creminelli:2006xe} and we refer the reader to those papers for a detailed explanation.
The construction of the effective theory is based on the following consideration. In a quasi de
Sitter background with only one relevant degree of freedom, there is a privileged spatial slicing
given by the physical clock which allows us to smoothly connect to a decelerated hot Big Bang
evolution. The slicing is usually realized by a time evolving scalar $\phi(t)$, but this does not necessarily need to be the case. To describe perturbations
around this solution one can choose a gauge where the privileged slicing coincides with surfaces of
constant $t$, i.e. $\delta\phi(\vec x,t)=0$. In this `unitary' gauge there are no explicit scalar perturbations but only metric
fluctuations. As time diffeomorphisms have been fixed and are not a gauge symmetry anymore,
the graviton now describes three degrees of freedom: the scalar perturbation has been eaten by the
metric. One therefore can build the most generic effective action with operators that are functions
of the metric fluctuations and that are invariant under the linearly-realized time-dependent spatial
diffeomorphisms. As usual with effective field theories, this can be done in a low energy expansion
in fluctuations of the fields and derivatives. We obtain the following Lagrangian \cite{Cheung:2007st,Creminelli:2006xe}:
 \begin{eqnarray}
\label{eq:actiontad}\nonumber
\!\!\!S  =\int  \! d^4 x \; \sqrt{- g} &\Big[&\frac12 M_{\rm 
Pl}^2 R + M_{\rm Pl}^2 \dot H
g^{00} - M_{\rm Pl}^2 (3 H^2 + \dot H)+ \\\nonumber 
&&+ \frac{1}{2!}M_2(t)^4(g^{00}+1)^2+\frac{1}{3!}M_3(t)^4 (g^{00}+1)^3+ \\
&& - \frac{\bar M_1(t)^3}{2} (g^{00}+1)\delta K^\mu {}_\mu
-\frac{\bar M_2(t)^2}{2} \delta K^\mu {}_\mu {}^2
-\frac{\bar M_3(t)^2}{2} \delta K^\mu {}_\nu \delta K^\nu {}_\mu + ...\, \Big],
\end{eqnarray}
where we denote by $\delta K_{\mu\nu}$ the variation of the extrinsic curvature of constant time surfaces with respect to the unperturbed FRW: $\delta K_{\mu\nu}=K_{\mu\nu}-a^2 H h_{\mu\nu}$ with $h_{\mu\nu}$ being the induced spatial metric, and where $M_{2,3}$ and $\bar M_{1,2,3}$ represent some time-dependent mass scales.

Let us comment briefly on (\ref{eq:actiontad}). The first term is the Eistein-Hilbert term. The first three terms are the only ones that start linearly in the metric fluctuations. The coefficients have been carefully chosen to ensure that when combined the linear terms in the fluctuations cancel. The action must start quadratic in the fluctuations. The terms in the second line start quadratic in the fluctuations and have no derivatives. The terms in third line represent higher derivative terms. Dots represent operators that start at higher order in the perturbations or in derivatives. This is the most general action for single field inflation and in fact it is unique~\cite{Cheung:2007st}.

The unitary gauge Lagrangian describes three degrees of freedom: the two graviton helicities and
a scalar mode. This mode will become explicit after one performs a broken time diffeomorphism
(St\"uckelberg trick) to reintroduce the Goldstone boson which non-linearly realizes this symmetry. In analogy
with the equivalence theorem for the longitudinal components of a massive gauge boson \cite{Cornwall:1974km}, the physics of the Goldstone decouples from the two graviton helicities at high enough energies, equivalently  the mixing can be neglected. The detailed study of \cite{Cheung:2007st,Senatore:2009gt} shows that in most situations of interest this is indeed the case and one can neglect the metric fluctuations\footnote{Equivalently, the neglected effects are suppressed by slow-roll parameters or by powers of $H/\mpl$. }. 

As anticipated, we reintroduce the Goldstone boson ($\pi$) by performing a broken time-diff., calling the parameter of the transformation $-\pi$, and then declaring $\pi$ to be a field that under time diff.s of the form $t\rightarrow t+\xi^0(x)$ transforms  as
\be
\pi(x)\quad\rightarrow\quad \tilde\pi(\tilde x(x))=\pi(x)-\xi^0(x)\ .
\ee
In this way diff. invariance is restored at all orders. For example the terms containing $g^{00}$ in the Lagrangian  give rise to the following terms:
\be
g^{00}\quad\rightarrow \quad\frac{\d (t+\pi)}{\d x^\mu}\frac{\d (t+\pi)}{\d x^\nu}g^{\mu\nu} \quad\rightarrow\quad g^{00} +2 g^{0\mu} \partial_\mu \pi + (\partial \pi)^2 .
\ee
We refer to~\cite{Cheung:2007st} for details about this procedure.
If we are interested just in effects that are not dominated by the mixing with gravity, then we can neglect the metric perturbations and just keep the $\pi$ fluctuations. In this regime, a term of the form $g^{00}$ in the unitary gauge Lagrangian becomes:
\be
g^{00}\quad\rightarrow\quad -1-2\dot\pi-\dot\pi^2+\frac{1}{a^2}(\d_i\pi)^2\ .
\ee
Further, we can assume that the $\pi$ has an approximate continuous shift symmetry, which becomes exact in the limit in which the space time is exactly de Sitter \cite{Cheung:2007st}. This allows us to neglect to terms in $\pi$ without a derivative that are generated by the time dependence of the coefficients in (\ref{eq:actiontad})\footnote{Notice that this is not always the case. Interesting inflation models, both single field and multifield, have been recently proposed in which the $\pi$ fluctuations are protected only by an approximate discrete shift symmetry. See for example \cite{Silverstein:2008sg,McAllister:2008hb,Green:2009ds,Barnaby:2009mc,Flauger:2009ab}.}. 
Implementing the above procedure in the Lagrangian of (\ref{eq:actiontad}), we obtain the rather simple result:
\begin{eqnarray}\label{eq:Spi}
S_{\rm \pi} =\int d^4 x   \sqrt{- g} \left[ -M^2_{\rm Pl}\dot{H} \left(\dot\pi^2-\frac{ (\partial_i \pi)^2}{a^2}\right)
+2 M^4_2
\left(\dot\pi^2+\dot{\pi}^3-\dot\pi\frac{(\partial_i\pi)^2}{a^2}
\right) -\frac{4}{3} M^4_3 \dot{\pi}^3+\ldots \right] \ ,
\eea
where for simplicity we have neglected the terms originating from the extrinsic curvature as they are usually important only in a regime where the space time is very close to de-Sitter space \cite{Cheung:2007st}. In fact, notice that neglecting the terms in the last line in (\ref{eq:actiontad}), we see that the spatial kinetic term is the form of $\dot H\mpl^2(\d_i\pi)^2$. In the limit $\dot H\rightarrow 0$ the spacetime approaches de Sitter and the coefficient of the spatial kinetic term goes to zero. If we consider the terms involving the extrinsic curvature in the last line of (\ref{eq:actiontad}), upon reintroduction of $\pi$ they give a spatial kinetic term of the form either $(\bar M_2^2+\bar M_3^2)(\d_i^2\pi)^2$ or $H\bar M_1^3 (\d_i\pi)^2$. In general these terms are either higher derivative terms or suppressed by the Hubble scale, and so they are in general  negligible at energies of order $H$. However, in the de Sitter limit the leading term $\dot H \mpl^2(\d_i\pi)^2$ goes to zero, and they become relevant. See \cite{Senatore:2009gt} for a precise definition of when this is the case. We refer to this situation as the near de Sitter limit and we will explicitly describe  the phenomenology in this limit at the end of the next section.

We notice that when $M_2$ is different from zero and we are not in the de Sitter limit the speed of sound of the fluctuations is different from one. We have the following relationship:
\be\label{eq:M2cs}
M_2^4=-\frac{1-c_s^2}{c_s^2}\frac{\mpl^2\dot H}{2}\ .
\ee
In this limit there are two independent cubic self interactions, $\dot\pi(\d_i\pi)^2$ and $\dot\pi^3$ at this order in derivatives, which can induce detectable non-Gaussianities in the primordial density perturbations. A small speed of sound (i.e. a large $M_2$) forces large self-interactions of the form $\dot\pi(\d_i\pi)^2$, while the coefficient of the operator $\dot\pi^3$ is not fixed because it depends also on $M_3$. By analysis of the cosmological data, one can therefore constrain (or measure) the parameters of the above Lagrangian. This approach has been recently applied to the WMAP data in \cite{Senatore:2009gt}, giving constraints on $M_2$ and $M_3$, as well as on the higher derivative operators that we have omitted in (\ref{eq:Spi}). This is the exact analogous of what happens for data from particle accelerators when the Precision Electroweak Tests of the Standard Model are carried out \cite{Peskin:1991sw,Barbieri:2004qk}.

\section{A detectable four-point function from single field inflation\label{sec:4-point-single-field}}

It is by now well established that single field inflation can produce a large and detectable level of 
non-Gaussianity in the three-point function. It is worth asking if it is also possible to have a large and detectable four-point function without at the same time having a detectable three-point function. The effective Lagrangian of single field inflation of sec.~\ref{sec:single-field} is the ideal general set up to address this kind of questions.

Restricting ourselves to the case where the Goldstone boson is protected by an approximate continuous shift symmetry, Ref.s~\cite{Cheung:2007st,Senatore:2009gt} show that in single field inflation there are only two ways to have a large three-point function: either by having a very small speed of sound $c_s$ for the fluctuations, or by the unperturbed solution being  so close to de Sitter space that the dispertion relation of the Goldstone boson is of the form $\omega^2\sim k^4/M^2$, where $M$ is some mass scale related to $\bar M_{2,3}$ and $M_2$. 

In the case of a small speed of sound and away from the de Sitter limit, the large three-point function is induced by the operators $\dot\pi(\d_i\pi)^2$ and $\dot\pi^3$ that are associated with the unitary gauge operators $(\delta g^{00})^2$ and $(\delta g^{00})^3$. In particular, by estimating loop corrections reference~\cite{Senatore:2009gt} showed that if the coefficient of the operator $(\delta g^{00})^2$ is $M_2^4\sim \dot H\mpl^2/c_s^2 $ using the $c_s\ll 1$ limit of (\ref{eq:M2cs}) then the operator $(\delta g^{00})^3$ is naturally of order $M_3^4\sim M_2^4/c_s^2$ (and viceversa). Both operators generate comparable three-point functions given by:\
\be
{ \langle \zeta^3\rangle \over \langle \zeta^2\rangle^{3/2}} \sim \left.\frac{L_3}{L_2}\right|_{E\sim H}\sim \frac{1}{c_s^2}\zeta \ ,
\ee
where to estimate the effect we have taken the ratio of the cubic and the quadratic Lagrangian at energies of order $H$. It is customary to define
\be
\langle\zeta_{\vec k_1}\zeta_{\vec k_2}\zeta_{\vec k_3 }\rangle=(2\pi)^3\delta^{(3)}(\vec k_1+\vec k_2+\vec k_3)(2\pi)^3\left(-\frac{9}{10}f_{NL} P_\zeta^2\right)\frac{1}{k^6}\,
\ee
where we have taken the equilateral limit of the 3-point function in Fourier space $|\vec k_1|=|\vec k_2|=|\vec k_3|=k$ and
\be
\langle\zeta_{\vec k_1}\zeta_{\vec k_2}\rangle=(2\pi^3)\delta^{(3)}(\vec k_1+\vec k_2)\frac{P_\zeta}{k^3}\ ,
\ee
so that
\be\label{eq:fnl}
f_{NL}\zeta \sim \left.\frac{{\cal L}_3}{{\cal L}_2}\right|_{E\sim H}\sim \frac{1}{c_s^2}\zeta \  \qquad\Rightarrow\qquad  f_{\rm NL}\sim\frac{1}{c_s^2}\ .
\ee

The Planck satellite is expected to reach a limit on $f_{NL}\sim$ few~\cite{Babich:2004yc} so we will limit ourselves to the case $c_s\ll1$. The reason why the natural value of $M_3^4$ is $M_2^4/c_s^2$ can be quickly understood by doing the following manipulations of the single field Lagrangian
\begin{eqnarray}\label{eq:Spics}
S_{\rm \pi} =\int d^4 x\,   \sqrt{- g} \left[ -\frac{M^2_{\rm Pl}\dot{H}}{c_s^2} \left(\dot\pi^2-c_s^2\frac{ (\partial_i \pi)^2}{a^2}\right)
-\frac{\mpl^2\dot H}{c_s^2}
\dot\pi\frac{(\partial_i\pi)^2}{a^2} -\frac{2}{3}  \frac{\tilde c_3}{c_s^4}\mpl^2\dot H \dot{\pi}^3 \right]\ ,
\eea
where we have used (\ref{eq:M2cs}), and we have redefined $M_3^4=\tilde c_3M_2^4/c_s^2$. We can perform a transformation of the spatial coordinates:
\be
\vec x\quad \rightarrow \quad \vec{ \tilde x}=\vec x/c_s\ ,
\ee
and canonically normalize the field $\pi$
\be
\pi_c=(-2\mpl^2 \dot H c_s)^{1/2}\;\pi\ ,
\ee
to obtain
\begin{eqnarray}\label{eq:Spicscan}
S_{\rm \pi}=\int dt\, d^3\tilde x\,   \sqrt{- g} &&\left[ \frac{1}{2} \left(\dot\pi_c^2-\frac{ (\tilde \partial_i \pi_c)^2}{a^2}\right)\right.\\ \nonumber
&&\left.-\frac{1}{\left(8|\dot H |\mpl^2c_s^5\right)^{1/2}}
\dot\pi_c\frac{(\tilde \partial_i\pi_c)^2}{a^2} -\frac{2}{3}  \frac{\tilde c_3}{\left(8|\dot H|\mpl^2c_s^5\right)^{1/2}} \dot\pi_c^3 \right]\ ,
\eea
where $\tilde \d_i=\d/\d\tilde x^i$. Notice that since we did not rescale time so the cutoff in energy can directly read off from (\ref{eq:Spicscan}). Since the kinetic part of the Lagrangian (\ref{eq:Spicscan}) is now Lorentz invariant, it is  easy to read off the unitarity bound
\be
 \Lambda^4\sim c_s^5|\dot H|\mpl^2 \sim c_s^7 M_2^4 \ .
\ee
In fact with this particular redefinition of the coefficient $M_3$ in terms of $\tilde c_3$ and $c_s^2$ we can see that for $\tilde c_3$ of order one the two operators $\dot\pi_c^3$ and $\dot\pi_c(\tilde \d_i\pi_c)^2$ are suppressed by the same scale $\Lambda^2$. Indeed it is  straightforward to verify that loop corrections will generate the two operators with their relative coefficient $\tilde c_3$ of order one. We further notice that having a detectable three-point function means lowering the cutoff of the theory well below $(\dot H\mpl^2)^{1/4}$. Note that in terms of the cut-off the size of the non-Gaussianity is
\be
{ \langle \zeta^3\rangle \over \langle \zeta^2\rangle^{3/2}} \sim \left({H \over \Lambda}\right)^2 \ ,
\ee
as in fact they are generated by dimension six operators.

Let us now consider the four-point function. The unitary-gauge operators $(\delta g^{00})^2$ and $(\delta g^{00})^3$ contain many quartic operators. For example:
\be
M_2^4(\d_i\pi)^4\ .
\ee
This operator induces a four-point function of the form
\be
{ \langle \zeta^4\rangle \over \langle \zeta^2\rangle^{2}} \sim \left.\frac{{\cal L}_4}{{\cal L}_2}\right|_{E\sim H}\sim \frac{1}{c_s^4}\zeta^2\ .
\ee
The four-point function is usually parametrized by $\tau_{\rm NL}= \langle \zeta^4\rangle / \langle \zeta^2\rangle^{3}$ so that,
\be\label{eq:taunl}
\tau_{\rm NL}\zeta^2\sim \left.\frac{{\cal L}_4}{{\cal L}_2}\right|_{E\sim H}\sim \frac{1}{c_s^4}\zeta^2 \qquad\Rightarrow\qquad  \tau_{\rm NL}\sim\frac{1}{c_s^4}\ .
\ee
This value of \taunl is tiny. The observational constraints errors on \fnl and \taunl scale as \cite{Creminelli:2006gc}
\be
\Delta [f_{NL} \zeta] \sim \frac{1}{N_{\rm pix}^{1/2}} \ ,\qquad \Delta[\tau_{NL} \zeta^2]\sim \frac{1}{N_{\rm pix}^{1/2}} \ ,
\ee
where $N_{\rm pix}$ represents the number of data points of the experiment.
In order for a four-point function to be detectable, the value of \taunl  has to be a factor of about  $10^5$ larger than the value of \fnl allowed by the data~\cite{Creminelli:2006gc}. Current limits from the WMAP satellite  \cite{Senatore:2009gt,Komatsu:2010fb} set $f_{NL}\lesssim 10^{2}$, which  from (\ref{eq:fnl}) implies that $c_s^2\gtrsim 10^{-2}$. Current limits on \taunl are therefore expected to be of the order $10^7$, and clearly the value in (\ref{eq:taunl}) is a factor of $10^3$ too small. Even a detection of \fnl at its current upper bound of order $10^2$ results in  $\tau_{NL}\sim 10^{4}-10^5$  which will not be detectable by the Planck satellite. Planck is expected to produce a limit on $\tau_{NL}$ of order $10^6$. In the future 21-cm line experiments that could map a large fraction of our Hubble volume might reach $f_{NL}\sim 10^{-2},\ \tau_{NL}\sim 10^3$ \cite{Loeb:2003ya}. In order to detect the four-point function coming from these operators we will have to wait for quite some time and it could only happen if there is a detection of a large three point function\footnote{If one looks carefully at the current best limits on the three-point function presented in~\cite{Senatore:2009gt}, one can notice that there is a small but non negligible region of the parameter space where values of $c_s$ as low as about $10^{-2}$ are allowed. This is due to a partial cancellation of the three-point function in that region of parameter space, which however is not expected to happen at the level of the four-point function. This therefore implies a value of $\tau_{NL}$ of order $10^8$ with such an high level of non-Gaussianity that could be detectable already by the WMAP satellite. Obviously, for this to happen we would have to be lucky enough so that inflation occured in that particular and  small region of parameter space where the three-point function is   suppressed.}. So far we have considered only the quartic operator $(\d_i\pi)^4$ induced by the unitary-gauge operators $(\delta g^{00})^2$ and $(\delta g^{00})^3$. It is straightforward to see that the situation is the same also for the other operators associated with them.

Notice that because of the non-linear realization of time-diffs, the coefficients of the quartic operators induced by $(\delta g^{00})^2$ and $(\delta g^{00})^3$ where tied to the ones of the cubic operators. This was the reason why the induced four-point function was so small. However, one could imagine introducing the operator $M_4^4(\delta g^{00})^4$ which starts quartic in the fluctuations and to arbitrarily set to zero the coefficients of the operators $(\delta g^{00})^2$ and $(\delta g^{00})^3$.The fact that the coefficient of $\dot\pi^4$ is unrelated to the coefficients of the cubic terms had been already noticed in \cite{Chen:2009bc} for a subclass of the models we consider consisting of scalar field Lagrangians of the form $P((\d\phi)^2,\phi)$. However, without the identification of a symmetry protecting the generation of cubic terms, it is hard to imagine why one should have concentrated on the particular Lagrangian allowing for a large quartic operator and small cubic ones. The point of our paper is that we are able to identify such a symmetry.

Reintroduction of the field $\pi$ would give rise to terms of the form
\be
M^4_4(\delta g^{00})^4\quad\rightarrow\quad M^4_4\left(16 \dot\pi^4-32 \dot\pi^3(\d_\mu\pi)^2+24 \dot\pi^2(\d_\mu\pi)^4-8\dot\pi(\d_\mu\pi)^6+(\d_\mu\pi)^8\right)\ ,
\ee
The term in $\dot\pi^4$ induces a four-point function whose size is given by:
\be\label{eq:taunl2}
{ \langle \zeta^4\rangle \over \langle \zeta^2\rangle^{2}} \sim \tau_{\rm NL}\zeta^2\sim \left.\frac{{\cal L}_4}{{\cal L}_2}\right|_{E\sim H}\sim \frac{M^4_4}{\dot H\mpl^2}\zeta^2 \qquad\Rightarrow\qquad  \tau_{\rm NL}\sim  \frac{M_4^4}{\dot H\mpl^2}\ \ .
\ee
Let us concentrate on the limit $\tau_{NL}\gg 1$ ($\dot H\mpl^2\gg \Lambda_U^4$) as this is the only regime where the signal is detectable. Upon canonical normalization, one can read that the unitarity bound of the theory is
\be
\Lambda_U^4\sim \frac{(\dot H \mpl^2)^2}{M^4_4}\sim \frac{(\dot H \mpl^2)}{\tau_{NL}}\ .
\ee
$\tau_{NL}\gg 1$ translates into the condition $\dot H\mpl^2\gg \Lambda_U^4$, and we can rewrite $\tau_{NL}\zeta^2$ as
\be
\tau_{NL}\zeta^2\sim\left.\frac{{\cal L}_4}{{\cal L}_2}\right|_{E\sim H}\sim\frac{H^4}{\Lambda_U^4}\ .
\ee
Thus by making the cutoff closer and closer to the Hubble scale we make the non-Gaussianities detectable.  However in taking this limit we should worry about the fact that we have arbitrarily set to zero the operators $(\delta g^{00})^2$ and $(\delta g^{00})^3$ which we expect to be generated at loop level. Let us see at what level they are generated. Let us consider two of the terms in $\pi$ included in the operators $(\delta g^{00})^4$: $\dot\pi^4$ and $\dot\pi^3(\d_\mu\pi)^2$. Upon canonical normalization, and neglecting factors of order one, we can see that these two operators appear in the Lagrangian as
\be
\frac{1}{\Lambda_U^4}\dot\pi_c^4 \ , \qquad \qquad \frac{1}{\Lambda_U^6\tau_{NL}^{1/2}}\dot\pi_c^3(\d_\mu\pi_c)^2 \ .
\ee
For $\tau_{NL}\gtrsim 1$, the operator $\dot\pi_c^3(\d_\mu\pi_c)^2$ is effectively suppressed by an higher scale than $\dot\pi_c^4$. The same holds for the higher dimension operators that are included in $(\delta g^{00})^4$: the higher the power in $\pi$ they have the higher is the scale by which they are suppressed. This theory has an approximate and accidental $\pi\rightarrow-\pi$  symmetry.

If we now ask at what level these operators will generate the operators $(\delta g^{00})^2$ and $(\delta g^{00})^3$ we realize that upon reinsertion of $\pi$ these operator do not respect the symmetry $\pi\rightarrow -\pi$: they contain even and odd powers of $\pi$. In fact their leading interaction term is odd.  To generate them we will have to insert the operator $\dot\pi_c^3(\d_\mu\pi_c)^2$, effectively suppressing the level at which the operators  $(\delta g^{00})^2$ and $(\delta g^{00})^3$ are generated. In other words, in the limit $\tau_{NL}\gtrsim 1$ there is an approximate $Z_2$ symmetry $\pi\rightarrow-\pi$ which is softly broken and protects the renormalization of the operators  $(\delta g^{00})^2$ and $(\delta g^{00})^3$. 

For example, it is easy to estimate that the loop in fig.~\ref{fig:3-point loop generation} induces an operator $\dot\pi^3_c$ suppressed by the scale $(\dot H\mpl^2)^{3/2}$. Since the structure of the Lagrangian is fixed by symmetries, by returning to the original normalization of the fields we conclude that  the operators  $(\delta g^{00})^2$ and $(\delta g^{00})^3$ are generated with a coefficient of order $\dot H\mpl^2$:
\be\label{eq:op4point}
M^4_4(\delta g^{00})^4 \qquad \rightarrow \qquad \dot H\mpl^2 (\delta g^{00})^2\ , \dot H\mpl^2 (\delta g^{00})^3 \ .
\ee

\begin{figure}
\begin{center}
\includegraphics[width=5cm,]{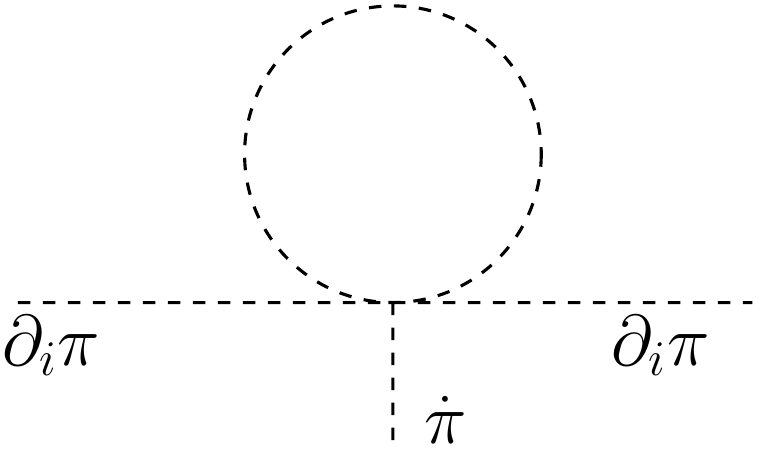}
\caption{\label{fig:3-point loop generation} \small Generation of a 3-point function from a loop 
diagram}
\end{center}
\end{figure}

This is a very interesting result. In fact doing the same estimates that lead to (\ref{eq:fnl}) we conclude that the induced value of $f_{NL}$ is just of order one, while at the same time the $\tau_{NL}$ in (\ref{eq:taunl2}) can be much larger than $10^5$. Therefore, we conclude that in single field inflation  we can have a four-point function that induces  a level of non-Gaussianities much larger than the three-point function and therefore can be detected even in the absence of a detection of a three-point function. Remarkably, this large four-point function is generated only by one operator: $\dot\pi^4$, and therefore there is a unique shape for it. Because it comes from derivative interactions this shape will be peaked on configurations in momentum space where all the momenta have comparable wavelengths.

In particular, $\tau_{NL}$ can be larger than $10^6$, which is the approximate threshold for WMAP. So this signal could be waiting to be discovered in existing WMAP data. The theory and analysis of a four-point function in the CMB data is a largely unexplored area.  We find that this result strongly motivates such a study. We will undertake it in a separate paper~\cite{kensen1}. 
 
Of course, in order to have  a large four-point function, it is not necessary that the operators $(\delta g^{00})^2$ and $(\delta g^{00})^3$ be exactly set to zero. It is  straightforward to see that in the case of speed of sound different from one the condition for having the four-point function be the leading non-Gaussianity reads
\be
M^4_4\gtrsim \frac{(\dot H\mpl^2)^{3/2}}{H^2 c_s^{7/2}}\ .
\ee

Finally, in models with a background space-time very close to de Sitter space \cite{Cheung:2007st,Senatore:2009gt} the operator $\dot\pi^4$ still gives a four-point function larger that the three-point function if the coefficient $M_4$ is large enough. There are however different possible shapes for the four-point function depending on the dispersion relation of modes with frequencies close to $H$.  In the case where this is of the form $\omega=\gamma^2 k^2/M_2$, with $\gamma^2=(\bar M_2^2+\bar M_3^2)/(2 M_2^2)$, the operators of the form $(\delta K^\mu{}_\mu)^2$ and $\delta K^\nu{}_{\mu}\delta K^\mu{}_\nu$ dominate at horizon crossing, and  the condition for having a four-point function larger than the three-point function reads
\be\label{constraints-4-point-ghost}
M^4_4\gtrsim \frac{M_2^{25/4}}{H^{9/4} \gamma^{1/2}}\ ,\quad M_4^4\gtrsim \frac{\bar M_1^3 M_2^{13/4}}{H^{9/4}\gamma^{5/2}}\ ,
\ee
where for simplicity we have taken $\gamma\lesssim 1$ in order not to have an effective speed of sound larger than one at energies below the cutoff.  This estimate follows from realizing that the wavefunction in Fourier space for the canonically normalized $\pi_c$ goes as 
\be
\pi_{c,k}\sim\frac{1}{\sqrt{\omega}}\sim\frac{M_2^{1/2}}{\gamma k}\ ,
\ee
which implies that at energies of order $E$, the size of the vacuum fluctuations of the operator $\pi(x)$ goes as
\be
\left.\pi\right|_E\sim\frac{1}{M_2^2} \left(\int d^3 k \frac{M_2}{\gamma^2 k^2}\right)^{1/2}\sim \frac{1}{M_2}\left(\frac{E}{M_2}\right)^{1/4}\frac{1}{\gamma^{3/2}}\ .
\ee
The rest follows by re-doing the estimates we did in the case of a dispersion relation of the form $\omega=c_s k$. Notice that in this case the curvature perturbation $\zeta$ is of order
\be
\zeta\sim H \left.\pi\right|_{E\sim H}\sim \left(\frac{H}{M}\right)^{5/4}\frac{1}{\gamma^{3/2}}\sim 10^{-5}\quad \Rightarrow \quad H\sim 10^{-4}\gamma^{6/5} M_2\ ,
\ee
and the first constraint in (\ref{constraints-4-point-ghost}) becomes
\be
M_4\gtrsim 10^2 \frac{M_2}{\gamma^{4/5}} \ . 
\ee
Notice also that the cubic operator in $M_2$ generates a three-point function of order 
\be
\left. \frac{{\cal L}_3}{{\cal L}_2}\right|_{E\sim H}\sim\left(\frac{H}{M_2}\right)^{1/4}\frac{1}{\gamma^{7/2}}\sim 10^{-1}\frac{1}{\gamma^{16/5}}\ .
\ee
Constraints on the three-point function~\cite{Senatore:2009gt,Komatsu:2010fb} imply $\gamma\gtrsim 5$. 
Even though the operator generating a four-point function is still $\dot\pi^4$ the wavefunctions are different in the two cases  so the induced shape of the four-point function in Fourier space will be a different. In practice the difference is not expected to be large, as is the case for the the three-point function~\cite{Senatore:2009gt}. On the other hand, it is easy to estimate that the operator $\dot\pi^4$ that in unitary gauge is associated with the operator $(\delta g^{00})^4$ will generate unitary-gauge operators of the form $\delta K^4$ and $(\delta g^{00})^2\delta K^2$, where for simplicity we have suppressed the spacetime indexes which can be contracted in all possible ways. These  operators will appear with coefficients  that make them as important as $(\delta g^{00})^4$. Notice that the scaling dimension for all of these operators is the same as the dispersion relation is $\omega\propto k^2$. Upon reinsertion of $\pi$, these operators induce couplings of the form $(\d^2_i\pi)^4$ and $\dot\pi^2(\d_i^2\pi)^2$ where the $\d_i$ are spatial derivatives and the indexes are contracted in all possible ways. The operators of the form $\delta g^{00}\delta K^3$ and $(\delta g^{00})^3\delta K$ (with spacetime indexes contracted in the several possible ways) are not generate in a relevant way because they violate the symmetry $\pi\rightarrow-\pi,\ t\rightarrow -t$. On the other hand, we could have started directly from these operators of the form $\delta g^{00}\delta K^3$ to find that all the former ones are generated with similar size. Summarizing, we see that we can have a total of twelve different quartic operators in $\pi$ involving either one time-derivative or two space derivatives acting on each $\pi$ with various possible tensorial contractions.  These twelve different operators have comparable shapes  and comparable sizes. This is only possible in the near de Sitter models with dispersion relation of the form $\omega=\gamma^2k^2/M_2$.

In the de Sitter limit and if the dispersion relation is of the form $\omega=c_s k$ ($c_s^2=H\bar M_1^3/(4M_2^4)$) the unitary gauge operator $\delta g^{00}\delta K^\mu{}_\mu$ sets the dispersion relation at horizon crossing. The condition for the four-point function induced by the operator $(\delta g^{00})^4$ to dominate reads:
\be\label{eq:constraindesitternew}
M^4_4\gtrsim \frac{M_2^6}{H^2 c_s^{1/2}}\ .
\ee
Notice that the three-point function in this case goes as $\zeta/c_s^2$~\cite{Senatore:2009gt}, where $\zeta$ is given by \cite{Senatore:2009gt}
\be
\zeta\sim\left(\frac{H}{M_2}\right)^2\frac{1}{c_s^{3/2}}\sim 10^{-5}\ .
\ee
The constraint in (\ref{eq:constraindesitternew}) becomes $M_4^4\gtrsim M_2^4/(\zeta c_s^2)$ in order for the four-point function to be the dominant signal. It is consistent to assume that the operators of the form  $(\delta K^\mu{}_\mu)^2$ and $\delta K^\nu{}_{\mu}\delta K^\mu{}_\nu$ that induce a dispersion relation of the form $\omega\propto k^2$ have a coefficient such that they are important only at the unitarity bound induced by the $\dot\pi^4$ operator. In this case the dispersion relation is linear in $k$ up to the cutoff and it is easy to see that the quartic operators involving the extrinsic curvature are now generated at a negligible level. In this case, the shape of the four-point function is just given by the operator $\dot\pi^4$ and its shape is identical to that in the not-near de Sitter models. If one were to include larger coefficients for the operators $(\delta K^\mu{}_\mu)^2$ and $\delta K^\nu{}_{\mu}\delta K^\mu{}_\nu$ then one would continuously interpolate to the case of a dispersion relation of the form $\omega\propto k^2$.


\section{Conclusions}

We have shown that in single-clock inflation it is possible to generate a four-point function that could be detected even in the absence of a detectable three-point function. This is possible by imposing an approximate continuos shift symmetry and an approximate parity symmetry in the Effective Lagrangian for the inflationary fluctuations. For models away from the near de Sitter limit, there is a {\it unique} interaction operator that gives rise to the large four-point function: $\dot\pi^4$. This results in only {\it one} possible shapes for the four-point function. Also in the models close to de Sitter  the four-point function can be the leading signal, and in this case there is more that one possible shape. This motivates us to undertake an analysis of the WMAP data in search for such a signal~\cite{kensen1}.

\subsubsection*{Acknowledgments}

We thank Nima Arkani-Hamed, Paolo Creminelli, Shamit Kachru and Eva Silverstein for interesting conversations. L.S.~is supported in part by the National Science Foundation under PHY-0503584.
M.Z. is supported by the David and Lucile 
Packard Foundation, the Alfred P.~Sloan Foundation and the John D. and Catherine T.~MacArthur Foundation.

 \begingroup\raggedright\endgroup

\end{document}